# Natural Way of Solving a Convex Hull Problem

Sina Saadati[1], Mohammadreza Razzazi[2]

## Abstract

In this article, a new solution for the convex hull problem has been presented. The convex hull is a widely known problem in computational geometry. As nature is a rich source of ideas in the field of algorithms, the solution has been inspired by nature. A tight elastic band is modeled using agents and also nails as points of the problem. By simulating an elastic band with nails in an environment, solving the convex hull problem will be possible. The algorithm runs in O(t) in which t is the time that an elastic band will get fixed.

## 1 Introduction

A convex hull S of a set of points in the plane is the smallest convex set that contains all the points of the set and for any pair of points Q1 and Q2 belonging to set S, the line segment Q1Q2 is completely contained in S. In other words, it is the intersection of all convex sets that contain S [1, 2, 3, 4, 8]. Imagine points being nails sticking out of the plane, taking an elastic rubber band, holding it around the nails, and letting it go. It will snap around the nails, reducing their length. The convex hull of P is the area surrounded by the rubber band. It results in an alternative definition of the convex hull of a set of finite points P in the plane: it is the special convex polygon whose vertices are points from P and include all points inside P [1, 2, 3].

The convex hull is one of the known problems in the field of computational geometry. Some algorithms solve it in different ways. Grham's Scan algorithm solves the convex hull problem in O(N logN) time. Jarvis's march algorithm is another method that is output-sensitive and solves the problem in O(NH) time where H is the number of the vertices in the plane that are included in the polygon of the convex hull. Convex hull can also be solved using divide-and-conquer in O(N logN) [1, 5, 6, 7, 9, 10].

In this paper, we are trying to discover how nature solves this problem and based on this natural solution, we will simulate a tight elastic band and a number of nails located on a surface using agent-based modeling.

This paper is organized as follows: Section 2 is about the reality of the movements of the band with an elasticity feature. In this section, we will study how the atoms and tiny particles of an elastic band behave in the nature based on physical theories. In Section 3, we will explain an agent-based algorithm that enables us to simulate an elastic band and use it to solve the convex hull problem.

## 2 Preliminaries

### 2.1 Newton's second law

Newton's second law explains the movement of objects which are affected by a force. According to this law, when an object is influenced by a force, it will be accelerated on the same side of the force. The value of acceleration depends on the value of the force and mass of the object. It is accepted that:

$$F = ma \quad (1)$$

In which F denotes the total force, m denotes the mass of the object and a denotes the acceleration of the object. In this equation, we assume F and a as vectors[11]. In physics, when two hard objects collide, the atoms of the two objects approach each other. Electrons are in the outermost parts of the atom and each electron has a negative electric field around it. The intensity of the field increases and atoms try to repel each other. This force increases as the distance between the two particles decreases and at one point it does not allow the atoms to get closer anymore. So, if we assume the environment as a discrete area of cells, we can claim that each cell can obtain at most one particle inside[11].

[1] Department of Computer Engineering, Amirkabir University of Technology, Tehran, Iran, Sina.Saadati@aut.ac.ir
[2] Department of Computer Engineering, Amirkabir University of Technology, Tehran, Iran, Razzazi@aut.ac.ir

## 2.2 Space is discrete

Confirmed evidence is used to believe that physical space-time is quantum, meaning that space can be assumed to be a set of cells located close to each other. In these methods, we will implement the space as an array of subareas where each subarea can include a maximum of one particle inside. By using this data model, we will look at the space around each particle in O(1) time that is similar to what is happening in the nature[12, 13].

## 2.3 Elastic Band

An elastic band is a flexible string with elasticity. The elasticity in this material is due to the gravity between the constituent particles of the band. For simplicity in this article, we consider the elastic band as a one-dimensional string of particles. It means each particle in this object is in relation to only two adjacent particles and it moves according to their gravity.

In figure 1, we have focused on one particle (object number 3). In this figure, the white particle is influenced by the gravitational force from two adjacent particles (objects numbers 1 and 2). The total force of T1 and T2 is shown by T3.

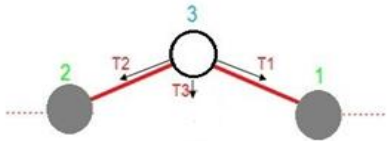

*Figure 1: Three particles of an elastic band which are in gravitational connection*

Therefore, the equation of motion for the particle number 3 can be expressed as[11]:

$$F = ma \qquad (2)$$

$$a = \frac{\partial V}{\partial t} \qquad (3)$$

$$V = \frac{\partial X}{\partial t} \qquad (4)$$

where F is the total force affecting an object, a is acceleration created by F, V is the velocity of the object and X is the location of the object. F, a, V and X are assumed as vectors in a 2D environment. Here t represents the time. As in agent-based modeling, time is assumed as a discrete variable, so we can rewrite the equations as follows[14]:

$$F = m \, \Delta a \qquad (5)$$

$$a = \frac{\Delta V}{\Delta t} \qquad (6)$$

$$V = \frac{\Delta X}{\Delta t} \qquad (7)$$

In this article, we consider the elastic band as a set of tiny particles. Then we implement each particle as an agent. There for, each particle can own its acceleration(a) and velocity(V) in every moment of the simulation. By having the value V as a vector, movement of particles can be simulated according to eq. (7).

The movement of a particle can be also affected by particles of nails. Each time we move any particle, we must check the area around it. Movement is possible only if there is nothing in its way.

We also must consider the friction between particles of the elastic band and the surface. This force must be considered when we want to compute the total force in eq. (2). Friction will affect the objects in a way that tries to stop them. So it can be claimed:

$$Friction = \mu N \qquad (8)$$

if the object is moving.

$$Friction = -Fs \qquad (9)$$

if the object is not moving, Yet it is affected by another force like Fs. (Fs can be equal to zero.)

In eq. (8) N is the vertical force pushing the object from the ground. It can be seen in figure2. $\mu$ is a constant value that depends on the type of surface.

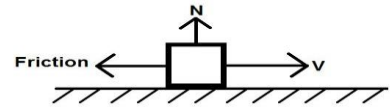

*Figure 2: Friction affecting a moving object.*

In our model, we can assume N as the reaction force to the gravity from the ground. So it is claimed:

$$|N| = mg \qquad (10)$$

Where g is equal to 9.8 and m is the mass value of the object. As m and g never change in our model, we can assume N as a constant value. As $\mu$ is constant too, we can assume dynamic friction as a force with constant value but on different sides [11].

# 3 Elastic Band Algorithm

After learning the behavior of particles affected by forces, we can implement an agent-based model of an elastic band. The implementation can be done by an object-oriented programming language like java or by an agent-based simulation tool like Repast. We have implemented our model using both methods, java programming language and repast[15].

In every agent-based modeling, we should consider how agents get synchronized with each other. In many projects(including ours) behavior of agents must be run in a predefined order. In this paper, first, we compute the state of each agent including their acceleration and velocity. After all agents clarified their states, we can move them. Otherwise, the movement of particles won't be correct. In the Java programming language, we use the algorithm as follows:

```
algorithm eachTime(){
Input: All agents of elastic band at time t.
Output: All agents at time t+1.
1.    for each particle in elastic band{
2.        particle.calculateState();
3.    }
4.    for each particle in elastic band{
5.        particle.move();
6.    }
   }
```

The algorithm eachTime must be called at each time step of our simulation. In repast, we run the same algorithm to keep the agents synchronized with each other. The algorithm below must be written inside each particle of the elastic band:

```
algorithm eachTime(){ \\for repast agents
Input: One agent of elastic band at time t.
Output: One agent of elastic band at time t + 1.
1.    static integer timeCounter = 0 ;
2.    timeCounter ++ ;
3.    if(timeCounter % 2 == 0 ) {
4.        this.calculateState();
5.    }else{
6.        this.move();
7.    }
  }
```

calculateState algorithm uses physical equations to compute the acceleration of the particle. Then it will update the speed according to the acceleration value. It has been written as follows:

```
algorithm calculateState(){
Input: An agent of elastic band with state of time t .
Output: An agent of elastic band with state of time  t + 1.
1. VectorA = Force from right neighbour particle of the agent
2. VectorB = Force from left neighbour particle of the agent.
3. TotalForce = VectorSum(VectorA , VectorB);
4. update speed using TotalForce and equations 5 and 6.
5. checkFriction();

}
```

We know that for computing the friction, we must check if the particle is moving or not. That is why we update the speed before computing the friction. The algorithm of computing friction can be defined as follows:

```
algorithm checkFriction(){
Input: An agent of elastic band which is not affected by friction
Output: An agent of elastic band affected by friction
1.  V = speed of the particle as a vector.
2.  if(  |V| > ε ){
3.      |V| = |V| − ε
4.  } else {
5.      |V| = 0;
6.  }
  }
```

Notice that V in the algorithm is a vector. So, we must do vector calculations in lines 3 and 5. After calculating the state of particles, we must move them. The movement can be done by using eq.7. Existence of a nail in front of a particle prevents its movement. Thus, every time a particle moves, it should look around and figure out the possibility of mevement. The algorithm of movement has been written as follows:

```
algorithm move(){
Input: Status of a particle  in the environment at time t.
Output: Status of the particle in the environment at time t+1.
1.   PointA = location of the particle at the time(t).
```

2. PointB = location of the particle at next time(t+1). It can be calculated by eq.7 and speed vector of the particle.
3. tempVector = Make a vector from PointA to PointB.
4. Reduce the length of tempVector.
   (Set $|tempVector| = \varepsilon$).
5. while (particle has not reached the PointB) {
      Move the particle towards tempVector if it is allowed. If there is a nail particle in the way, set $|V| = 0$. and break while cycle.
7. }
   }

The environment in this project been implemented as a 2D array. Each cell of the array was considered as a subarea. Particles arrive in a subarea only if the subarea is empty. For simplicity, in this paper, we assume that particles of the elastic band are very small so that the collision between particles of the elastic band never happens. Output of implementation of our method are shown in figures 3 to 5. Figures 3, 4, and 5 show three steps of the process of solving the problem. In figure 3, we have positioned the elastic band around the set of nails. Here, we can imagine that we are keeping the elastic band with four fingers. In figure 4 we have removed the fingers and the band is getting tighter. In figure 5 we see that the elastic band is as tight as it can be and it represents the convex hull.

## 4  Conclusion

In this article, we have implemented an agent-based algorithm which behaves similar to an elastic band in the nature. Then we have discussed how to solve the convex hull problem using this method.

We have implemented the method using the Java programming language and repast. Agents were implemented to play the role of a particle. Then, we used the program to implement an elastic band and a set of nails. By having these tools, we are able to solve the convex hull problem for each set of points.

During the project, we understood that there is a computational difference between our model in java and repast. For searching the nearby grid cells, our model in java takes O(1) time while the model in repast takes much more time. In java, we used a 2-dimension array for the environment. In java, every entry of the array represents a rigid cell of the environment, so we have applied our algorithm to search the cells. In repast, we used the predefined functions of its libraries for the search of cells, which may be the reason for the overhead.

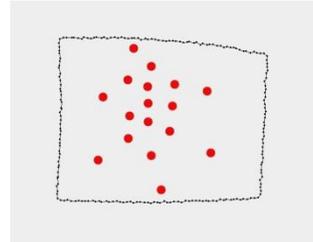

*Figure 3: The elastic band is located by four fingers.*

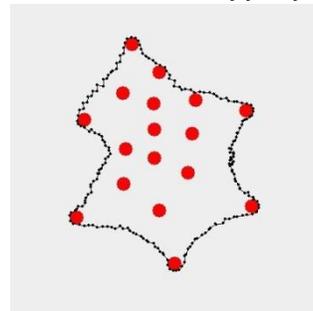

*Figure 4: The elastic band is getting tighter.*

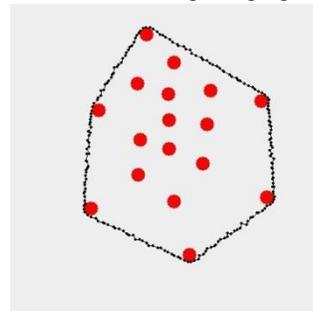

*Figure 5: The elastic band is showing the convex hull.*

# 5 Declarations

## 5.1 Funding

Not applicable.

## 5.2 Conflicts of interest/Competing interests:

Not applicable.

## 5.3 Availability of data and material:

Not applicable

## 5.4 Code Availability

You can access the developed models at:
https://github.com/sinasaadati95/elastic_band_models
In this link, there are two folders. First one is our developed model using Java programming language. The second one is the developed model using RepastSimphony.

## 5.5 Author's Contributions

The idea of solving geometric problems by imitation of nature has been suggested by second author and first author helped to model an elastic band and its implementation with Java and Repast.

# 6 References

[1]  M.D. Berg and O. Cheong and M.V. Kreveld and M. Overmars, Computational Geometry, Algorithms and applications. 3rd ed., Berlin, Heidelberg: SpringerVerlag, 2008.

[2]  A.M. Andrew, Another efficient algorithm for convex hulls in two dimensions, Information Processing Letters., vol. 9, pp.216-219, 1979.

[3]  G. S. Brodal and R. Jacob, "Dynamic planar convex hull," The 43rd Annual IEEE Symposium on Foundations of Computer Science, 2002. Proceedings., 2002, pp. 617-626.

[4]  A. Byakat, Convex hull of a finite set of points in two dimensions, Information Processing Letters, vol. 7, pp. 95-108, 1978.

[5]  T.M. Chan, Output-sensitive results on convex hulls, extreme points, and related problems, Discrete & Computational Geometry, vol. 16, pp. 369-387, 1996.

[6]  T.M. Chan, Dynamic planar convex hull operations in near-logarithmic amortized time, Journal of the ACM (JACM), vol. 48, pp. 1-12, 2001.

[7]  B. Chazelle, An optimal convex hull algorithm and new results on cuttings, in Proceedings 32nd Annual Symposium of Foundations of Computer Science, 1991, pp. 29-38.

[8]  J.E. Goodman and J.O. Rourke, Handbook of discrete and computational geometry., CRC press, 2017.

[9]  M.R. Kabat, Design and analysis of algorithms, 1st ed., PHI Learning Pvt. Ltd., 2013.

[10]  S.L. Devadoss and J. O'Rouke, Discrete and Computational Geometry., Princeton University Press, 2011.

[11]  D. Halliday and R. Resnick and J. Walker, Fundamentals of Physics. 10th ed., John Wiley & Sons, 2013.

[12]  S. Holzner, Physics For Dummies., John Wiley & Sons, 2010.

[13]  G.C. Gibson, Creation and Cosmos - The Literal Values of Genesis., Lulu Press Inc., 2017.

[14]  A.V. Oppenheim and A.S. Willsky and S. Hamid, Signals and systems, Processing series., Prentice Hall Upper Saddle River, 1997.

[15]  E. Bonabeau, "Agent-based modeling: Methods and techniques for simulating human systems," Proceedings of the National Academy of Sciences, vol. 99, no. Supplement 3, pp. 7280–7287, May 2002.